\documentclass[%
 reprint, superscriptaddress, 
 amsmath,amssymb, aps, prr, 
 longbibliography
 ]{revtex4-2}
\usepackage{graphicx}
\usepackage{subfigure}
\usepackage{multirow}
\usepackage{comment}
\usepackage[usenames]{color}
\usepackage{dcolumn}
\usepackage{bm}
\usepackage[colorlinks=true, allcolors=blue]{hyperref}
\usepackage{amsmath}
\usepackage{mathtools}
\usepackage{xcolor}
\usepackage{dcolumn}
\usepackage{bm}
\usepackage[normalem]{ulem}
\usepackage{lipsum}
\usepackage{float}

\graphicspath{ {./Figures/} }

\newcommand{\NoteR}[1]{{\color{red} (RPS: {#1})}}

\begin{document}
\title{Symmetry breaking in collective decision-making  through higher-order interactions}

\author{David March-Pons}
\email{david.march@upc.edu}
\author{Romualdo Pastor-Satorras}
\affiliation{Departament de Física, Universitat Politècnica de Catalunya, Campus Nord B4, 08034 Barcelona, Spain}

\author{M. Carmen Miguel}
\affiliation{Departament de Física de la Matèria Condensada, Universitat de Barcelona, Martí i Franquès 1, 08028 Barcelona, Spain.}
\affiliation{Institute of Complex Systems (UBICS), Universitat de Barcelona, Barcelona, Spain
}

\begin{abstract}
    
    Collective decision-making is a widespread phenomenon in both biological and artificial systems, where individuals reach a consensus through social interactions. While traditional models of opinion dynamics and contagion focus on pairwise interactions, recent research emphasizes the importance of including higher-order group interactions and autonomous behavior to better reflect real-world complexity. 
    In this work, we introduce a collective decision-making model inspired by social insects. In our framework, uncommitted agents can explore options independently and become committed, while social interactions influence these agents to prefer options already accepted by the group. Our model extends classical contagion models by incorporating multiple, mutually exclusive options and distinguishing between pairwise and higher-order social influences. Using simulations and analytical mean-field solutions, we show that higher-order interactions are essential for breaking symmetry in systems with equally valid options. We find that pairwise communication alone can cause decision deadlock, but adding group interactions allows the system to overcome stalemates and reach consensus. Our results emphasize the important roles of autonomous behavior and higher-order structures in collective decision-making. These insights could help us better understand social systems and design decision protocols for artificial swarms.
    

\end{abstract}

\maketitle

\section{Introduction}

Collective decision-making is a self-organized phenomenon by which a group of agents,  each with their own preferences and social environment, reaches a consensus through interactions within a social network~\cite{gerritsUnderstanding2017,living_in_groups_krause,baronchelli_emergence_2018,bose2017,Jackson2010}. This process is observed across a wide range of biological systems—from humans and social mammal herds to schooling fish, social insects, and even artificial robotic swarms~\cite{dyer2009, living_in_groups_krause, collective_animal_beh_book, animal_signals_book, valentini2017, Leonard2024}.
Typically, these processes involve choosing between at least two competing, mutually exclusive options, although the decision spectrum can often be viewed as continuous—such as in collective movement~\cite{vicsekCollectiveMotion2012} or polarization on specific topics~\cite{OjerExplosive2023}.
Many models have been proposed to understand how consensus forms, starting with simple rules like imitation in the voter model~\cite{castellano_statistical_2009} or the majority rule model~\cite{galam2002}. Numerous extensions have been developed to incorporate features such as individual heterogeneity, the influence of personalized information, or nonlinear interactions~\cite{CastellanoPRE2009, galam2008, redner_reality-inspired_2019, de_marzo_emergence_2020}.

Concurrently, contagion models—used to describe phenomena like the spread of diseases or rumors—have been extensively studied~\cite{pastor_epidemic_2001, pastor_epidemic_2015, daley_rumors_1964, dearruda_2018}. Similar to opinion-dynamics models, the basic approach assumes simple pairwise interactions within networks, which is sufficient to capture straightforward scenarios like epidemic spread. However, this approach falls short in modeling more complex behaviors. For instance, empirical evidence shows that multiple exposures to a source are often necessary to trigger a change in social behavior~\cite{centola2010, rosenthal2015}. This has led to the development of complex contagion models, such as the threshold model~\cite{granovetter1978threshold, watts2002simple, watts_influentials_2007}. Nonetheless, these models still assume that contagion occurs solely through pairwise interactions mediated by social contacts.
Since social interactions are in many cases group-based and not pairwise~\cite{starnini_face_2013, starnini_face_2016, gallo_face_ho_2024, musciotto2022_animals_ho}, higher-order structures such as hypergraphs or simplicial complexes have been proposed as more appropriate models for these interactions~\cite{battiston2020, Bianconi_2021, Battiston2021}.
This has led to a significant body of research on how contagions spread over such higher-order structures in different social settings~\cite{iacopini_simplicial_2019, deArruda_contagion_hg_2020, barrat_inbook_2022, FerrazdeArruda_contagion_2024, iacopini_tete_2024}.
Building on this, recent efforts have focused on uncovering the new phenomena of opinion dynamics and competitive spreading processes that occur on higher-order structures~\cite{Sahasrabuddhe_2021, noonan_majRule_hg_2021, kim_voter_hg_2025, zhao_comp_sirs_2024, li_competing_2022, iacopini_ng_2022}.

Here we study a collective decision-making model where agents can autonomously commit to available options. This framework is inspired by the behavior of social insects like honeybees, which communicate to share information about different tasks such as house-hunting or foraging, while also exploring and choosing options independently~\cite{Seeley2006amSci, seeley_stop_2012}. As a result, these models resemble epidemic spreading models, but with the key difference that multiple contagions spread simultaneously over the system~\cite{pais2013, reina_desing_pattern, reina2017, gray_multiagent_2018, Leonard2024}.
As a novelty, we propose a collective decision-making model in which pairwise communication and higher-order (group) interactions act as independent channels of information spreading. This extends the social contagion model studied in Ref.~\cite{iacopini_simplicial_2019} to account for multiple, mutually exclusive opinions
~\cite{li_competing_2022, nie_markovian_2022}.
Furthermore, in this work, we focus on the effects of the individual adoption component, which—though often overlooked in epidemic spreading models—is a key factor in collective decision-making, encompassing both individual exploration and social exchange of information.


Our findings highlight the relevance of group interactions in symmetry breaking scenarios, where the system must choose between equivalent options. 
In such cases, pairwise recruitment interactions are not enough for the system to make a decision, often leading to a stalemate~\cite{reina2017, marchpons2024consensus}. The inclusion of cross-inhibition, a negative signaling mechanism by which agents dissuade their peers holding contrary opinions from their current state, has been shown to be necessary for the system to make a decision in this contex~\cite{seeley_stop_2012, reina_cross_inhibition_2023, marchpons2025_nlinCI}. Cross-inhibition involves opinionated agents simultaneously recruiting neutral or uncommitted agents while dissuading peers with opposing views. Conversely, in systems lacking this negative inter-opinion signaling—if they only involve communication between opinionated and neutral agents—group interactions can serve as an alternative symmetry-breaking mechanism.

The paper is organized as follows. In Sec.~\ref{seq:model_def}, we define the collective decision-making model under study. In Sec. ~\ref{seq:model_mf}, we present a mean-field analysis, identifying stable deadlock states, symmetry-breaking solutions, and the nature and location of the transitions. Finally, in Sec.~\ref{seq:simulationsRSC} and Sec.~\ref{seq:simulationsRealSC}, we validate the mean-field predictions by simulating the model on synthetic random simplicial complexes and on complexes generated from real datasets, respectively.

\section{Model definition}
\label{seq:model_def}

The underlying structure in which our model of social contagion unfolds is that of a simplicial complex~\cite{Bianconi_2021}. 
Differently from classic binary networks~\cite{Newman10}, this representation is suited to describe higher-order interactions, in which groups of different sizes (from the usual pairwise connections to triangles, squares, etc.),  affect differently the spreading of the contagion processes.
Briefly, a simplicial complex is a subtype of hypergraph in which all links of smaller dimension (subsimplices) contained in a simplex of dimension $D$ are also considered~\cite{battiston2020}; e.g. given a full triangle ($D=2$) formed by vertices $(i,j,k)$ each pairwise link (of dimension $D=1$) $(i,j)$, $(i,k)$, $(j,k)$ is also considered as an open spreading channel.

To model the spread of multiple options on a simplicial complex, we define the state variables $s_i(t)$, representing the state of agent $i$ at time $t$. The state $s_i(t) = 0$ indicates a neutral or uncommitted agent, while $s_i(t) = 1, ..., M$ denotes an agent that holds and spreads one of the $M$ possible options $\alpha = 1, ..., M$. Similar to epidemiological models, these states represent whether an agent is susceptible or infectious regarding the spread of the options in the system. This model can be viewed as an extension of the Simplicial Contagion Model introduced in Ref.~\cite{iacopini_simplicial_2019}, which features the simultaneous spreading of multiple social contagions. 

In our framework, agents committed to option $\alpha$ recruit neutral peers at a rate $\beta_o^{\alpha}$, where $o = 1, ..., D$ indicates the order of the interaction, i.e. the size of the group that is informing an uncommitted agent—assuming all members of the group share the same state $\alpha$.
Thus, $\beta_1^\alpha$ indicates the usual pairwise contagion through a link $(i,j)$, where $s_i = 0$ and $s_j = \alpha$; $\beta_2^{\alpha}$ indicates the contagion through a full triangle $(i,j,k)$, where $s_i = 0$ and $s_j = s_k = \alpha$, and higher-order terms follow the same pattern.

At the same time, agents can spontaneously adopt opinion $\alpha$ independently of social influence. This autonomous transition occurs at a rate $\nu_{\alpha}$. This socially unfiltered change captures processes such as agents autonomously exploring and adopting options—similar to house-hunting insects~\cite{seeley_stop_2012, reina2017}—or interactions driven by individual-specific factors~\cite{de_marzo_emergence_2020}.

Finally, committed agents abandon their state and return to the uncommitted state at a rate $r_{\alpha}$. This mechanism allows agents to stop spreading their option and reassess their state, based either on individual exploration or the influence of their surroundings. In many collective decision-making models, this rate is often taken to be inversely proportional to the quality or benefit associated with option
$\alpha$~\cite{reina_desing_pattern, reina2017}. This creates a positive feedback loop in which higher-value options are reported to neutral agents for longer periods, increasing the likelihood that the group eventually converges on one of these options.

\section{Mean Field solution}
\label{seq:model_mf}


In this work, we focus on the spreading and competition of $M = 2$ equivalent options. Consequently, the rates of spontaneous adoption, recovery, and recruitment are assumed to be equal for both options and can be considered as global parameters of the system: $\nu$, $r$ and $\beta_o$.
To characterize the behavior of the model, we first analyze it under the mean field approximation, assuming homogeneous mixing~\cite{iacopini_simplicial_2019,kiss2017mathematics}. 
The temporal evolution of the population holding state $i$ is thus described by the  rate equation
\begin{equation}
    \frac{d x_i (t)}{dt} = \nu x_0(t) - r x_i(t) + \sum_{o = 1}^D \beta_o \langle k_o \rangle x_i^o(t) x_0(t)
    \label{eq:general_hoi_xi}
\end{equation}
where $x_0(t)$ is the fraction of uncommitted individuals, given by $x_0(t) = 1 - \sum_{j=1}^M x_j(t)$. The parameter $ \langle k_o \rangle $ represents the average simplicial degree of a 0-dimensional face (node), i.e., the average number of links, triangles, and higher-order simplices incident on a particular node. In this study, we will only consider simplices up to dimension $D = 2$, meaning spreading occurs via pairwise links and full triangles. By rescaling time with $r$ and defining the rescaled parameters $\pi = \nu/ r$, $\lambda_1 = \beta_1 \langle k \rangle / r$, and $\lambda_2 = \beta_2 \langle k_2 \rangle / r$, where $k \equiv k_1$ is the usual node degree, we arrive at the final rate equation for each $x_i$
\begin{equation}
     \frac{d x_i (t)}{dt} = \pi x_0(t) - x_i(t) + [\lambda_1 x_i (t) + \lambda_2  x_i^2(t) ] x_0(t).
\end{equation}
In the case $D=2$, 
it is more convenient to work with two alternative variables: the density of active population, $\rho(t)$, and the magnetization or consensus parameter $m(t)$~\cite{marchpons2024consensus,seeley_stop_2012}, defined as
\begin{equation}
    \rho(t) = x_1(t) + x_2(t) , \;\;m(t) = x_1(t) - x_2(t),
\end{equation}
respectively. Using these variables, the rate equations for the binary option case take the final form

\begin{eqnarray}
    \dot{\rho} &=& (1-\rho) \left[ 2 \pi + \lambda_1 \rho + \frac{\lambda_2}{2} (\rho^2 + m^2) \right] - \rho  \label{eq:rho_dot}\\
    \dot{m} &=& [(1-\rho)(\lambda_1 + \lambda_2 \rho) - 1]m.
    \label{eq:m_dot}
\end{eqnarray}

From this set of equations,  we can determine the steady state solutions $(\rho^*, m^*)$ by imposing $\dot{\rho} = \dot{m} = 0$. This allows us to identify the transition from a deadlock state ($m^* = 0$)—where no option is preferred—to a symmetry-breaking phase ($m^* \neq 0$), in which one option gains the lead. 

\subsection{Deadlock solution}
\label{seq:deadlock_sol}

By imposing  $m^*=0$ in the steady state expression of Eq.~\eqref{eq:rho_dot}, we obtain the polynomial equation
\begin{equation}
    -\frac{\lambda_2}{2} \rho^3 + \left( \frac{\lambda_2}{2} - \lambda_1 \right) \rho^2 + (\lambda_1 - 2\pi - 1) \rho + 2 \pi = 0 \text{.}
    \label{eq:deadlock_rho_poly}
\end{equation}
When there is no spontaneous adoption, i.e. when $\pi = 0$, the trivial solution  $\rho^*_d = 0$ exists and is stable for $\lambda_1 \leq 1$. Additionally, a pair of non-zero solutions emerge from the resulting second-order equation, taking the following form
\begin{equation}
    \rho_{d,\pm}^* (\pi = 0) = \frac{1}{2} - \frac{\lambda_1}{\lambda_2} \pm \frac{1}{\lambda_2} \sqrt{\left( \frac{\lambda_2}{2} - \lambda_1 \right)^2 + 2 \lambda_2 (\lambda_1 - 1)} \text{.}
\end{equation}
For $\lambda_1 \leq 1$, the non-zero solutions are positive and real only if $\lambda_2 \geq 4 - 2\lambda_1 + 4\sqrt{1-\lambda_1}$. In this case, $\rho_{d,+}^*$ is a saddle, while $\rho_{d,-}^*$ is unstable. When $\lambda_2 < 4 - 2\lambda_1 - 4\sqrt{1 - \lambda_1}$, both solutions become negative and thus are not physically relevant. 
If $\lambda_1 > 1$, only $\rho_{d,+}^*$ yields a positive real solution and acts as a saddle, while $\rho_{d,-}^* < 0$. Therefore, for $\pi = 0$, the only stable and physically meaningful solution is the trivial absorbing state $\rho^*_d = 0$  when $\lambda_1 \leq 1$.

For $\pi \neq 0$, Eq.~\eqref{eq:deadlock_rho_poly} can be solved analytically using Cardano's formula~\cite{uspensky1948theory}. Since these solutions are highly cumbersome, we instead adopt a numerical approach. In this case, we find one real solution ($\rho_d^* > 0$) and a pair of complex conjugate solutions. The stability of the real solution can be inferred from the stability of the symmetry-breaking solution discussed next.

\subsection{Symmetry breaking solution}
\label{seq:symbreking_sol}

\begin{figure}[t!]
    \centering
    \includegraphics[width=0.9\columnwidth]{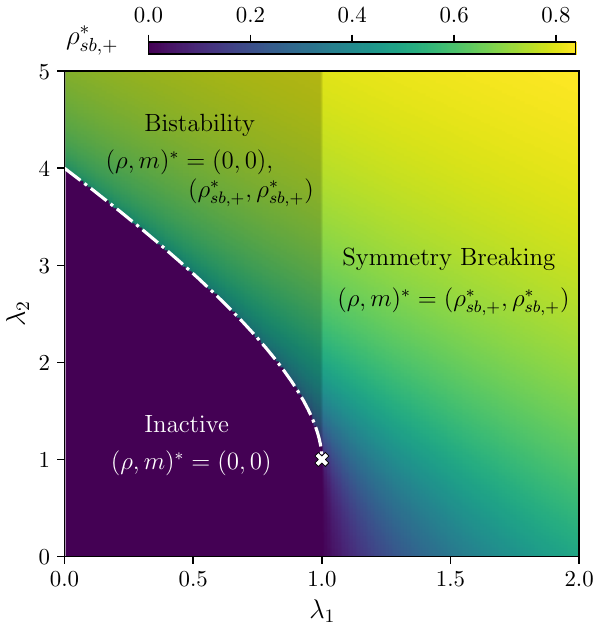}
    \caption{Mean field value of the active population, $\rho_{sb,+}^*$, for the case of two contagions spreading without spontaneous adoption ($\pi = 0$) in the parameter space defined by the pairwise and three-wise spreading intensities, $(\lambda_1, \lambda_2)$. The consensus state is given by $|m^*| = \rho_{sb,+}^*$. The dot-dashed line indicates the discontinuous transition at $\lambda_1^{c,\pi = 0} = 2\sqrt{\lambda_2}-\lambda_2$, while the shaded area delineates the bistability region.}
    \label{fig:fig1con}
\end{figure}

Imposing a finite value of $m^* > 0$ in the steady state of Eq.~\eqref{eq:m_dot} leads to  the equation
\begin{equation}
    (1-\rho)(\lambda_1 + \lambda_2 \rho) = 1,
    \label{eq:poly_rho_sb}
\end{equation}
which yields the solutions
\begin{equation}
    \rho_{sb, \pm}^* = \frac{\lambda_2 - \lambda_1 \pm \sqrt{(\lambda_1 - \lambda_2)^2 - 4 \lambda_2 (1 - \lambda_1)}}{2 \lambda_2}\text{.}
    \label{eq:rho_sb}
\end{equation}
These solutions are analogous to those reported in  Refs.~\cite{iacopini_simplicial_2019,li_competing_2022} for the cases of one or two competing epidemic processes, respectively. 
Notably, while the active population in the stationary state does not depend on the spontaneous adoption rate $\pi$, the parameter $m$ does. By substituting Eq.~\eqref{eq:poly_rho_sb} into the steady state of Eq.~\eqref{eq:rho_dot}, we obtain
\begin{equation}
    m^*_{sb} = \pm \sqrt{\rho_{sb}^{*2} - \frac{4 \pi }{\lambda_2}} \text{.}
    \label{eq:m_sb}
\end{equation}
This result implies that when $\pi = 0$, eventually one option dominates completely while the other vanishes, leading to a consensus $m^*_{sb} = \pm \rho_{sb}^*$. Although our focus is on a binary decision problem, it can be shown that this ``one takes it all'' solution is indeed the only valid and stable outcome for a general decision among $M$ options when spontaneous adoption is absent.

\begin{figure*}[t!]
    \centering
    \includegraphics[width=0.9\linewidth]{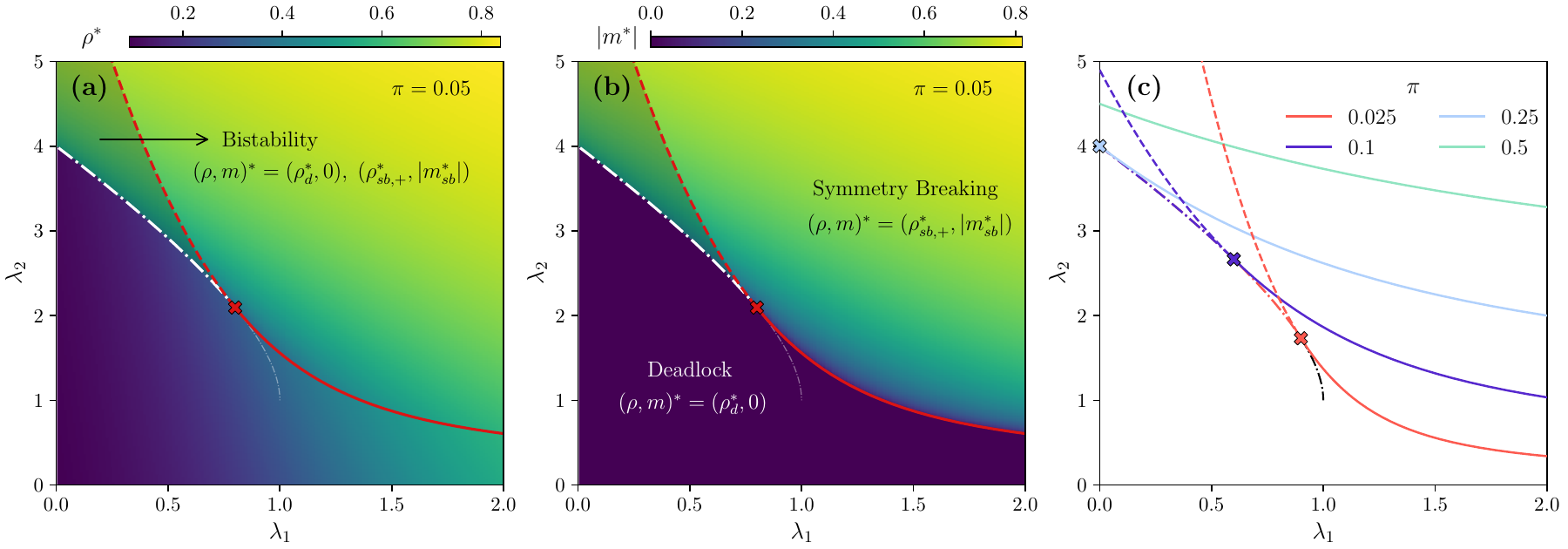}
    \caption{Mean field results for two contagion spreading with spontaneous adoption, $\pi = 0.05$, in the $(\lambda_1, \lambda_2)$ parameter space. \textbf{(a) and (b)}: The heatmaps show the solutions for $\rho^*$ and $m^*$, respectively. The dot-dashed line indicates the discontinuous transition at $\lambda_1^c = 2\sqrt{\lambda_2}-\lambda_2$. The continuous line marks the continuous transition, which shifts from $\lambda_1 = 1$ when $\pi  = 0$ to the values given by Eq.~\eqref{eq:B2_lines}($+$). The dashed line, given by Eq.~\eqref{eq:B2_lines}($-$), marks the boundary of the bistability between the deadlock and symmetry-breaking solutions. The black cross indicates the point $(\lambda_1^{\times},\lambda_2^{\times})$ (Eq.~\eqref{eq:l1x_l2x}), where all the threshold lines intersect and the nature of the transition changes. In both figures, the different regions are labeled according to their valid solutions, with the shaded area indicating the bistability region. Specifically, this region is colored based on the symmetry-breaking solution.
    \textbf{(c)}: Effect of the spontaneous adoption parameter $\pi$ on the threshold lines. When  $\pi < 1/4$, the continuous line (Eq.~\eqref{eq:B2_lines}($+$)) indicates where the continuous transition occurs, while the dashed line (Eq.~\eqref{eq:B2_lines}(-)) indicates the upper boundary of the bistability region, which appears at the intersection point $(\lambda_1^{\times},\lambda_2^{\times})$, denoted by the cross markers.
    Note that the lower bound of the bistability region, $\lambda_1^c$, is common for all $\pi< 1/4$. }
    \label{fig:fig2con}
\end{figure*}

From Eq.~\eqref{eq:m_sb}, we observe that the spontaneous adoption rate $\pi$ acts as an effective  noise that hampers the final consensus: larger $\pi$ values reduce the value of $m_{sb}^*$, which measures the overall difference between the population committed to the winning and losing options. Therefore, the ideal scenario in terms of decision strength is setting $\pi = 0$. However, in this case, the system cannot incorporate external information into the decision process, which must start with some population already committed to an option. As highlighted in other works, strategies such as time-varying~\cite{talamali_improving_2019, talamali2021_less_more} or highly non-linear~\cite{Leonard2024, marchpons2025_nlinCI} signaling patterns are effective strategies to reduce the influence of spontaneous adoption as the decision process unfolds, thereby enhancing the likelihood of reaching a final consensus.

\subsubsection{Symmetry breaking without spontaneous adoption ($\pi=0$)}

The validity range of these solutions is determined by the square roots appearing in Eqs.~\eqref{eq:rho_sb} and~\eqref{eq:m_sb}. 
When $\pi = 0$,  we have $m^*_{sb} = \pm\rho_{sb}^*$, so only the validity of Eq.~\eqref{eq:rho_sb} needs to be considered, resulting in the same thresholds reported in~\cite{iacopini_simplicial_2019}: for $\lambda_2 \leq 1$, there is a continuous transition from $(\rho, m)^* = (0, 0)$ to $(\rho, m)^* = (\rho_{sb,+}^*, \pm\rho_{sb,+}^*)$ at $\lambda_1 = 1$ (one option vanishes and the other dominates the entire active population). In contrast, when $\lambda_2 > 1$, the transition becomes discontinuous and occurs at a threshold
\begin{equation}
    \lambda_1^{c, \pi = 0} = 2\sqrt{\lambda_2} - \lambda_2,
    \label{eq:B1_line}
\end{equation} 
which marks a sudden jump to the symmetry-broken state. While the null solution remains stable for
$\lambda_1 \leq 1$, there exists a bistable region between $\lambda_1^{c, \pi = 0}$ and $\lambda_1 = 1$, in which both the null and symmetry-breaking solutions are stable. Note that since the symmetry-breaking solution corresponds to consensus for either of the two options, this region can also be considered a tristability region, involving coexistence of three stable states: the null state and the two possible consensus states.  

Ultimately, three regions---inactive, bistable and symmetry breaking---appear in the  $(\lambda_1, \lambda_2)$ parameter space, as shown in Fig.~\ref{fig:fig1con}. The tricritical point at $(\lambda_1, \lambda_2) = (1, 1)$ (marked by a white cross) indicates where the nature of the transition shifts from continuous to discontinuous. Notably, when the system lacks group interactions ($\lambda_2 = 0$), the mean-field analysis reduces exactly to the contact process, which exhibits the critical behavior of the directed percolation universality class~\cite{marchpons2024consensus}.
Finally, the system also features non-stable solutions not shown in Fig.~\ref{fig:fig1con}: $\rho_{sb,-}^*$ is a saddle in the bistability region, and the physically irrelevant $\rho_{sb,-}^* < 0$ when $\lambda_1 > 1$. 

Additionally, the background of Fig.~\ref{fig:fig1con} shows the stationary values of $\rho_{sb,+}^*$, illustrating how the final density of the active population—and thus the degree of consensus—increases as either the pairwise or triadic communication intensity is enhanced. 

\subsubsection{Symmetry breaking with spontaneous adoption ($\pi > 0$)}


When $\pi > 0$, the situation becomes more nuanced. Although the symmetry-breaking solution for the active population may become real-valued, the square root in the consensus expression (Eq.~\eqref{eq:m_sb}) must also be considered. Substituting the active population solution Eq.~\eqref{eq:rho_sb} into the radicand of the consensus equation, we obtain the conditions that must be satisfied in order for real-valued symmetry-breaking solutions to exist:
\begin{equation}
    \begin{split}
        \pm (\lambda_2 - \lambda_1) \sqrt{(\lambda_2 + \lambda_1)^2 - 4\lambda_1} \; + \qquad \qquad \\  \; + \; \lambda_2^2 + \lambda_1^2 - 4\lambda_2 \geq 8 \lambda_2 \pi \text{.}
    \end{split}
    \label{eq:B2_lines}
\end{equation}

In this case, analytical expressions for the threshold lines cannot be obtained; instead, these expressions must be solved numerically for different values of $\pi$. The $\pm$ sign in Eq.~\eqref{eq:B2_lines} derives from the two solutions $\rho_{sb,\pm}^*$ in Eq.~\eqref{eq:rho_sb}. Overall, Eq.~\eqref{eq:B2_lines} defines two new threshold lines that complement the existing threshold $\lambda_1^{c, \pi = 0}$ (Eq.~\eqref{eq:B1_line}), replacing the continuous transition previously located at $\lambda_1 = 1$ and delimiting the three aforementioned regions anew -- see Figs.~\ref{fig:fig2con}(a,b). These new threshold lines intersect the $\lambda_1^{c, \pi = 0}$ line at a new tricritical point $(\lambda_1^{\times}, \lambda_2^{\times})$. This point can be precisely determined by substituting  Eq.~\eqref{eq:B1_line} into Eq.~\eqref{eq:B2_lines}(+), which yields
\begin{equation}
    \lambda_2^{\times} = 1 + 4(\pi + \sqrt{\pi}), \quad \lambda_1^{\times} = -\lambda_2^{\times}+2\sqrt{\lambda_2^{\times}} \text{.}
    \label{eq:l1x_l2x}
\end{equation}

The nature of the transition and specifically, which sign from Eq.~\eqref{eq:B2_lines} determines the threshold line, depends on whether $\lambda_2$ is smaller than or larger than $\lambda_2^{\times}$. When $\lambda_2 \leq \lambda_2^{\times}$, the system exhibits a continuous transition along the line given by Eq.~\eqref{eq:B2_lines}($+$) from the deadlock state $(\rho_d^*, 0)$, as described by Eq.~\eqref{eq:deadlock_rho_poly}, to the symmetry-breaking state $(\rho_{sb,+}^*, m^*_{sb})$, defined by Eqs.~\eqref{eq:rho_sb} and ~\eqref{eq:m_sb} -- see the continuous red lines in Figs.~\ref{fig:fig2con}(a,b).
Due to the presence of spontaneous adoption, this continuous transition shifts from $\lambda_1 = 1$ toward larger $\lambda_1$ values as $\lambda_2$ decreases. Therefore, the combination of spontaneous adoption and low pairwise spreading strength impairs the system’s ability to reach consensus. Additionally, the continuous transition regime extends beyond $\lambda_2 = 1$, since  Eq.~\eqref{eq:l1x_l2x} shows that $\lambda_2^{\times} > 1$ for any $\pi > 0$.

When $\lambda_2 > \lambda_2^{\times}$, it is Eq.~\eqref{eq:B2_lines}(-) that delimits a region of bistability between the deadlock state $(\rho_d^*, 0)$ and the symmetry breaking states $(\rho^*_{sb,+},m^*)$ -- see the white dot-dashed and red dashed lines in Figs.~\ref{fig:fig2con}(a,b). Similar to the $\pi = 0$ case, on increasing $\lambda_1$ the bistability region appears at $\lambda_1^{c, \pi = 0}$ (Eq.~\eqref{eq:B1_line}). However, in this scenario, it does not terminate at $\lambda_1 = 1$ but rather at a particular point given by Eq.~\eqref{eq:B2_lines}(-).
The presence of spontaneous adoption thus facilitates reaching the symmetry-breaking phase compared to the $\pi = 0$ case, as it reduces the size of the bistability region. This advantage comes at the cost of a smaller overall consensus value, since a fraction of the population remains committed to the losing options in the stationary state. 
Regarding the $\rho_{sb,-}^*$ solution, as in the $\pi = 0$ case, the state $(\rho_{sb,-}^*,m_{sb}^*)$ is a saddle in the bistability region, while $\rho_{sb,-}^* < 0$ in the symmetry breaking region. 

In the heatmaps of Figs.~\ref{fig:fig2con}(a,b), we show the values of $\rho^*$ and $m^*$ across the $(\lambda_1,\lambda_2)$ parameter space. As in the no spontaneous adoption scenario, both spreading intensities contribute to increasing the active population and the final consensus. Notably, in the spontaneous adoption case, $\rho^* > 0$, the continuous transition smoothly connects the deadlock and symmetry-breaking solutions, while the discontinuous transition in the bistability region manifests as a jump from $\rho_d^*$ to $\rho_{sb,+}^*$. Conversely, the consensus parameter goes from $m^*= 0$ to $m^*_{sb}>0$ either continuously or discontinuously (see that in Figs.~\ref{fig:fig2con}(a,b), the bistability region is colored according to the symmetry breaking solution).

Finally, in Fig.~\ref{fig:fig2con}(c), 
we show the effect of $\pi$ on the threshold lines. Larger values of $\pi$ increase  $\lambda_2^{\times}$, thereby enlarging the range of $\lambda_2$ values over which a continuous transition from deadlock to symmetry breaking occurs. However, the transition shifts to larger $\lambda_1$ as $\pi$ increases, making consensus more difficult to attain -- an effect similar to that reported in other studies lacking three-body communication~\cite{marchpons2024_kilobots, marchpons2024consensus}. Additionally, high values of $\pi$ result in a contraction of the bistability region, which facilitates reaching the symmetry-breaking state once $\lambda_2^{\times}$ is exceeded.
This trend persists up to $\pi = 1/4$, at which $(\lambda_1^{\times},\lambda_2^{\times}) = (0, 4)$, and the bistability region disappears. At this specific value of $\pi$, the system can break the symmetry between the two options even without pairwise communication. For $\pi > 1/4$, the system will display only a continuous transition between the deadlock and symmetry-breaking states.

\begin{figure*}
    \centering
    \includegraphics[width=0.96\linewidth]{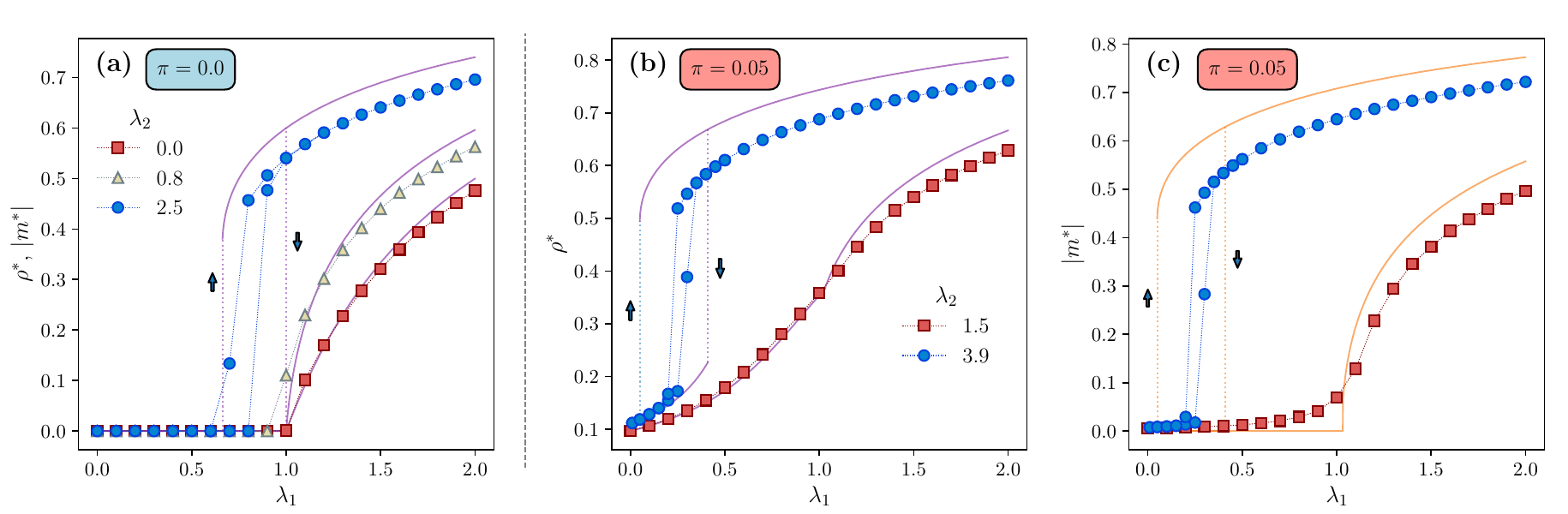}
    \caption{Simulations on a Random Simplicial Complex (RSC) with $\langle k \rangle = 20$ and $\langle k_2 \rangle = 6$. \textbf{(a)} No spontaneous adoption scenario, $\pi = 0$: three different values of the three-body spreading strength are shown ($\lambda_2 = 0$, red squares; $\lambda_2 = 0.8$, beige triangles; $\lambda_2 =2.5$, blue circles), while varying the two-body spreading strength $\lambda_1$. Since the simulations agree with the mean field results, we have $|m^*| = \rho^*$. \textbf{(b) and (c)} depict the spontaneous adoption scenario with $\pi = 0.05$. Two different values of the three-body spreading strength are shown ($\lambda_2 = 1.5$, red squares; $\lambda_2 = 3.9$, blue circles). Other simulation parameters are $r = 0.05$, $N = 2000$. }
    \label{fig:simsRSC}
\end{figure*}

\begin{figure*}
    \centering
    \includegraphics[width=0.96\linewidth]{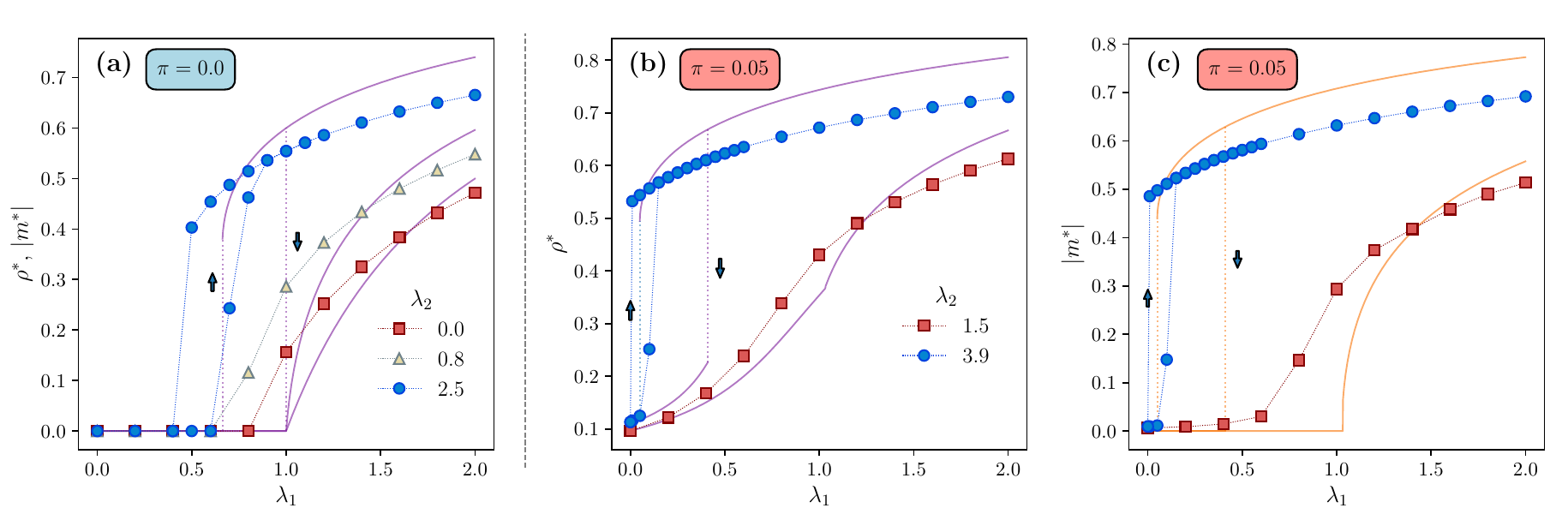}
    \caption{Simulations on a Simplicial Complex built from a real dataset based on time-aggregated face-to-face interactions in a high school. We use the same parameters as in the RSC simulations. \textbf{(a)}: The active population $\rho^* = |m^*|$ with no spontaneous adoption $\pi = 0$ and group recruitment strengths $\lambda =0.0, \; 0.8, \; 2.5$. \textbf{(b) and (c)} show $\rho^*$ and $|m^*|$, respectively, for spontaneous adoption $\pi = 0.05$ and group recruitment strengths $\lambda = 1.5, \; 3.9$. The system size of the simplicial complexes generated from the data is $N = 1611$, with an average node degree $\langle k_1 \rangle = 33.2$ and an average simplicial degree $\langle k_2 \rangle = 10.9$. 
    }
    \label{fig:simsThiers13}
\end{figure*}

\section{Simulations on Random Simplicial Complexes}
\label{seq:simulationsRSC}

To test the predictions of our mean-field model, we perform simulations on random simplicial complexes using the Random Simplicial Complexes (RSC) model~\cite{iacopini_simplicial_2019}. This model allows us to maintain a fixed average node degree $\langle k \rangle$ while varying the average number of triangles $\langle k_2 \rangle$ that a node participates in. 
Briefly, the RSC model of dimension $D$ allows for simplices of up to dimension $D$ and is defined by the number of nodes $N$ and the probabilities $\{p_1, ..., p_D\}$ for generating links (or simplices) of each dimension. Since we consider simplices up to dimension $D=2$, the model involves three parameters: $\{N, p_1, p_2\}$. The process begins by generating an Erdös-Rényi network by connecting nodes $i$ and $j$ with probability $p_1$. Next, triangles (2-simplices) are added by randomly selecting triplets of nodes $(i,j,k)$ with probability $p_2$. As a result, the RSC model generates simplicial complexes containing both empty triangles—3-cliques without three-body interactions—and full triangles—proper 2-simplices where three-body interactions are explicitly modeled. From Ref.~\cite{iacopini_simplicial_2019}, these probabilities are:
\begin{equation}
    p_1 = \frac{\langle k \rangle - 2 \langle k_2 \rangle}{(N-1) - 2 \langle k_2 \rangle}, \qquad p_2 = \frac{2 \langle k_2 \rangle}{(N-1)(N-2)}.
\end{equation}

We perform simulations of our multi-contagion process on RSCs composed of $N=2000$ nodes, with $\langle k_1 \rangle = 20$ and $\langle k_2 \rangle = 6$. For each run, a different RSC instance is generated, and nodes holding different opinions are assigned randomly. We then compute averages over the stationary state across 100 simulations for each set of parameters. In Fig.~\ref{fig:simsRSC}, we present the results for simulation sets conducted in the $\pi=0$ and $\pi>0$ scenarios.

In the $\pi = 0$ case shown in Fig.~\ref{fig:simsRSC}(a), the simulations show excellent agreement with the mean-field predictions, displaying a continuous transition for $\lambda_2 < 1$ and a discontinuous transition for $\lambda_2 > 1$. For $\lambda_2 = 2.5$, we have performed two sets of simulations with different initial conditions around the bistability region—one starting from a close-to null state and increasing $\lambda_1$, and the other starting from a close-to consensus state and decreasing $\lambda_1$—to capture both branches of the hysteresis cycle.
In accordance with the theoretical analysis, once the stationary state is reached, either both options vanish ($m^* = \rho^*_d = 0$) or one option dominates, occupying the entire active population ($m^* = \rho_{sb,+}^*$). In these cases, the active population and consensus are effectively the same and are represented in the same plot. Additionally, we consider the absolute value of the consensus, $|m^*|$, regardless of which option ultimately prevails.

In Fig.~\ref{fig:simsRSC}(b,c), we present simulation results for the same RSC properties ($N=2000$, $\langle k \rangle = 20$, $\langle k_2 \rangle = 6$) in the spontaneous adoption regime with $\pi = 0.05$. Subplots (b) and (c) show $\rho^*$ and $|m^*|$, respectively.
Similar to the previous case, the simulations exhibit good agreement with the mean-field theory, although the hysteresis cycle is narrower. To capture both branches of the cycle, we performed simulations starting from different initial conditions: since the final state of any population has $x_i > 0$, we used the final state from the previous simulation run---at a larger or smaller $\lambda_1$, depending on the branch— as the initial condition for the next simulation. In Supplemental Figure Fig.~SF1~\cite{suppmat}, we show the size dependence of the hysteresis cycle, indicating that its width increases as the system size grows.

The small-world arrangement of interactions in the random simplicial complex, combined with its strong connectivity, results in generally good agreement between the simulations and the mean-field predictions. However, when considering more sparsely connected random simplicial complexes, the deviations become more pronounced, both in the stationary values and the transition points—see
Supplemental Figure. Fig.~SF2~\cite{suppmat} for simulations on RSCs with degree $\langle k \rangle = 6$ and simplicial degree $\langle k_2 \rangle = 3$. Despite these discrepancies, the simulations still exhibit the same critical behavior for both $\pi = 0$ and $\pi = 0.05$.


\section{Simulation on other Simplicial Topologies}
\label{seq:simulationsRealSC}


To assess the mean-field predictions in other topologies with less controlled properties than those of the Random Simplicial Complexes, we use empirical data collected by the SocioPatterns collaboration~\cite{socipatterns}. These datasets consist of face-to-face interactions recorded in various social contexts, such as workplaces, high schools, and conferences. While they document only pairwise encounters, the time-resolved nature of the data allows us to construct group interactions by aggregating encounters occurring within short time windows.

More specifically, networks are generated by aggregating data within 5-minute time windows. From these, the maximal cliques are identified to create simplicial edges, and we retain the 20\% most frequent cliques of size 2 and 3 (as well as the size 2 and 3 subcliques of larger cliques, which are discarded). The data used for the simulations follow this methodology and are provided in~\cite{iacopini_simplicial_2019}. Moreover, to increase the size of the complexes, we employ the approach described in~\cite{young2017_sampling_scm}, which allows us to generate larger simplicial complexes while preserving the degree properties of the original data.

The simplicial complexes derived from real datasets differ from RSCs in that their generalized degree distributions are not centered around a single mean value; additionally, different social contexts yield different average generalized degrees. Nonetheless, simulations performed on these structures also reproduce the phenomenology predicted by the mean-field theory—and similarly observed in RSCs—as illustrated in Fig.~\ref{fig:simsThiers13} for a simplicial complex built from time aggregated face-to-face interactions in a high school~\cite{contacts_thiers13}. The average stationary values match the mean-field predictions closely, although the transition points occur at lower $\lambda_1$ values, primarily due to finite-size effects. In Supplemental Figure Fig.~SF3~\cite{suppmat}, we present results from simulations on a different substrate, constructed from interactions collected in a workplace~\cite{contacts_InVS15}. 
In this case, the agreement between simulations and mean-field predictions arises from the strong connectivity and effective well-mixing that result from the procedure used to construct the simplicial complexes.


\section{Discussion}

We have introduced a collective decision-making model based on the spread of mutually exclusive options through social interactions of varying group sizes. Recent studies highlight that group interactions are essential for accurately modeling social systems, as agents interact across a spectrum ranging from face-to-face encounters to large gatherings. Therefore, incorporating higher-order structures and group interactions is an essential step toward developing a more realistic analysis of opinion dynamics~\cite{Battiston2021}.

Inspired by models of collective behavior in social insects~\cite{reina2017, gray_multiagent_2018, Leonard2024}, our framework considers uncommitted agents being influenced by committed agents who attempt to recruit them. This approach builds on behavioral social contagion models—where a single state spreads through the system~\cite{iacopini_simplicial_2019, FerrazdeArruda_contagion_2024}—but now extends to the competition between two mutually exclusive options. 
A common feature of these models is the inclusion of autonomous adoption, representing individuals' personal quest for information. When combined with social recruitment, this mechanism plays a key role in determining whether and when the system can achieve consensus.

In this study, we focus on a collective decision between two equally beneficial alternatives—that is, options that provide the same benefit to the group. As shown in previous research, in such scenarios, a system that relies solely on pairwise interactions cannot break symmetry or make a definitive choice for one option~\cite{reina2017, marchpons2024consensus}. Other signaling mechanisms, such as cross-inhibition, have been proposed to enable the system to reach consensus~\cite{seeley_stop_2012, reina_cross_inhibition_2023, marchpons2025_nlinCI}. Here, we highlight the role of recruitment interactions between opinionated and neutral agents, demonstrating that extending to—and combining with—group interactions allows the system to regain the ability to break symmetry between the alternatives.

We have performed a mean field analysis to characterize the full parameter space defined by the strengths of pairwise and group (specifically, three-body) recruitment interactions. When there is no spontaneous adoption, only one contagion prevails, consistent with the phenomenology of the Simplicial Contagion Model~\cite{iacopini_simplicial_2019}. When spontaneous adoption is included, the competition between the two options persists; however, the presence of group interactions enables the system to transition from a deadlock to a symmetry-breaking state, where one option is held by a large majority of the population. 
Finally, we have tested the mean-field predictions on both synthetic structures (Random Simplicial Complexes) and simplicial complexes constructed from empirical datasets. The strong agreement with the mean-field results underscores the effectiveness of mean-field interactions in capturing the essential dynamics within these complex structures. This highlights the robustness of our analysis across diverse scenarios.

Ultimately, this study demonstrates that incorporating group recruitment interactions serves as an effective mechanism for breaking symmetry between equally beneficial alternatives. Our results highlight the crucial role of higher-order interactions in shaping consensus formation. Additionally, these findings open new avenues for designing distributed decision-making protocols in artificial swarms~\cite{valentini2017, reina_desing_pattern, Leonard2024}, where group interactions have yet to be fully explored. Moving forward, a key challenge is to investigate how this spreading model operates within structures based on face-to-face interactions—potentially dynamic in time—thus bridging the gap between the theoretical framework and its practical implementation in real-world systems.

\begin{acknowledgments}
  We acknowledge financial support from projects
  PID2022-137505NB-C21/AEI/10.13039/501100011033/FEDER\_UE, and
  PID2022-137505NB-C22 funded by MICIU/AEI/10.13039/501100011033, and by
  “ERDF: A way of making Europe”. 
\end{acknowledgments}


%



\end{document}